\documentclass[mathleft,fleqn]{an}
\usepackage{array}
\usepackage{hyperref}
\usepackage{natbib}
\usepackage{graphicx}
\usepackage[varg]{txfonts}
\usepackage{natbib}
\bibpunct{(}{)}{;}{a}{}{,}
\begin{document}

\Pagespan{1}{}
\Yearpublication{2017}%
\Yearsubmission{2017}%
\Month{0}%
\Volume{999}%
\Issue{0}%
\DOI{asna.201400000}%

\title{Supernova Ejecta in Ocean Cores Used as Time Constraints for Nearby Stellar Groups}

\author{Megan Hyde\inst{1}\fnmsep\thanks{Corresponding author:
        {hydem22@outlook.com}}
\and  Mark J. Pecaut\inst{1}
}
\titlerunning{Birth Site of Recent Supernova}
\authorrunning{M. Hyde \& M. J. Pecaut}
\institute{Department of Physics, Rockhurst University, 1100 Rockhurst Rd., Kansas City MO, 64110, USA}

\received{XXXX}
\accepted{XXXX}
\publonline{XXXX}

\keywords{Galaxy: open clusters and associations: individual (Scorpius-Centaurus, Tucana-Horologium) -- Methods: Data Analysis -- Stars: Evolution -- Stars: Statistics -- Stars: Supernovae}

\abstract{%
  Evidence of a supernova event, discussed in Wallner et al., was discovered in the deep-sea crusts with two signals dating back to 2-3 and 7-9 Myr ago. In this contribution, we place constraints on the birth-site of the supernova progenitors from the ejecta timeline, the initial mass function, and the ages of nearby stellar groups.  We investigated the Scorpius-Centaurus OB Association, the nearest site of recent massive star formation, and the moving group Tucana-Horologium.  Using the known stellar mass of the remaining massive stars within these subgroups and factoring in travel time for the ejecta, we have constrained the ages and masses of the supernova progenitors by using the initial mass function and then compared the results to the canonical ages of each subgroup.  Our results identify the Upper Scorpius and Lower Centaurus-Crux subgroups as unlikely birth-sites for these supernovae.  We find that Tucana-Horologium is the likely birth-site of the supernova 7-9 Myr ago and Upper Centaurus-Lupus is the likely birth-site for the supernova 2-3 Myr ago.}

\maketitle

\section{Introduction}
A supernova (SN) is an explosion caused by the death of a massive star that is at least 8$\pm$1$M_{\odot}$ \citep{smartt2009}.  SNe appear to have an upper limit of $\simeq$18$M_{\odot}$, where anything more massive collapses into a black hole \citep{smartt2015}. These SNe undergo extremely violent explosions that occur at a rate of $\sim$1-3 per century in our Galaxy and are the source of heavy metals in the Universe \citep{adams2013,fry2015}.  The recent discovery of the $^{60}$Fe signature in the oceanic crust has provided physical evidence for a near-Earth SN explosion \citep{fields2005,wallner2016}.  This physical evidence corresponds with the confirmation of a peak of SNe activity between 12-17 Myr within 200 pc of the solar system \citep{sorensen2017}.  A likely progenitor birth-site for this event is the Scorpius-Centaurus OB Association (Sco-Cen) because of the distance and the amplitude of the $^{60}$Fe \citep{benitez2002,breitschwerdt2016}. The goal of this contribution is to determine the birth-site of the SN that is consistent with the Initial Mass Function (IMF) and then compare the results to the mean ages of nearby stellar associations. 

Sco-Cen has historically been divided into three sub-groups:  Lower Centaurus-Crux (LCC), Upper Centaurus-Lupus (UCL), and Upper Scorpius (US). Sco-Cen is the nearest OB Association to the sun and is the nearest site of recent massive star formation \citep{preibisch2008}.  In this present work, each of these groups will be evaluated as possible points of origin and the results will be compared to their assumed mean ages.  It has also been suggested that Tucana-Horologium (Tuc-Hor) may have been responsible for the event \citep{mamajek2016}, so this group will also be considered.  

Smaller associations, such as the $\beta$ Pictoris Moving Group \citep{torres2008} or the 32 Ori Association \citep{bell2017}, will not be considered here since they are not massive enough to be the birth-site for a supernova progenitor. In addition, though evidence supports the existence of a larger young population in the vicinity of Taurus \citep{kraus2017}, the stellar census is incomplete and thus it will also be excluded in this study. 

\section{Data and Models}
\subsection{Models}
In order to establish a mass-lifetime relation, the rotating evolutionary tracks of masses from 1-70$M_{\odot}$ from the \citet{ekstrom2012} study were used. For an upper limit, 70$M_{\odot}$ was chosen since it is more than 3 times as massive than the largest star currently in Sco-Cen (see \autoref{tbl:USmass}). These tracks provide an accurate description of non-interacting stars and, including rotation, have been shown to accurately and simultaneously predict the main sequence width as well as the surface velocities and abundances \citep{ekstrom2012}.  These tracks are based on a grid of stars between 0.8-120$M_{\odot}$.  These stellar tracks adopted a metallicity of Z = 0.014 and include the enhancement of mass loss in the red supergiant stage which accurately predicts models above 15-20$M_{\odot}$ \citep{ekstrom2012}.  This accuracy is especially important because this investigation focuses on stellar masses that will eventually explode as a SN, so a reliable lifetime for each mass is needed in order to determine a possible point of origin for the SNe events discussed in \citet{wallner2016}. \par

Young associations within the solar neighborhood have near-solar metallicities \citep{almeida2009,barenfeld2013}, therefore, a solar metallicity of Z = 0.014 was adopted.  A rotational velocity of $V/V_{crit}$ = 0.355 was assumed because it is an average equatorial rotation, $<v_{eq}>$ = 120\,kms$^{-1}$, for the massive stars in the subgroups in Sco-Cen \citep{pecaut2012}. \par

Tuc-Hor is the largest young moving group of stars nearby, making it an excellent SN progenitor host association, with a distance of $\sim$40pc away \citep{zuckerman2001,kraus2014}. Moving groups are believed to have come from a dispersed yet `coeval' population \citep{zuckerman2004,mamajek2001}, meaning they will have similar characteristics such as metallicity and distance.  These young moving groups have near-solar metallicities \citep{almeida2009}, so, for this project, it will be assumed that Tuc-Hor has the same metallicity and rotational velocities as Sco-Cen.  \par

\subsection{Mass of Associations}
\begin{figure}[h!]
    \centering
    \includegraphics[width=\columnwidth]{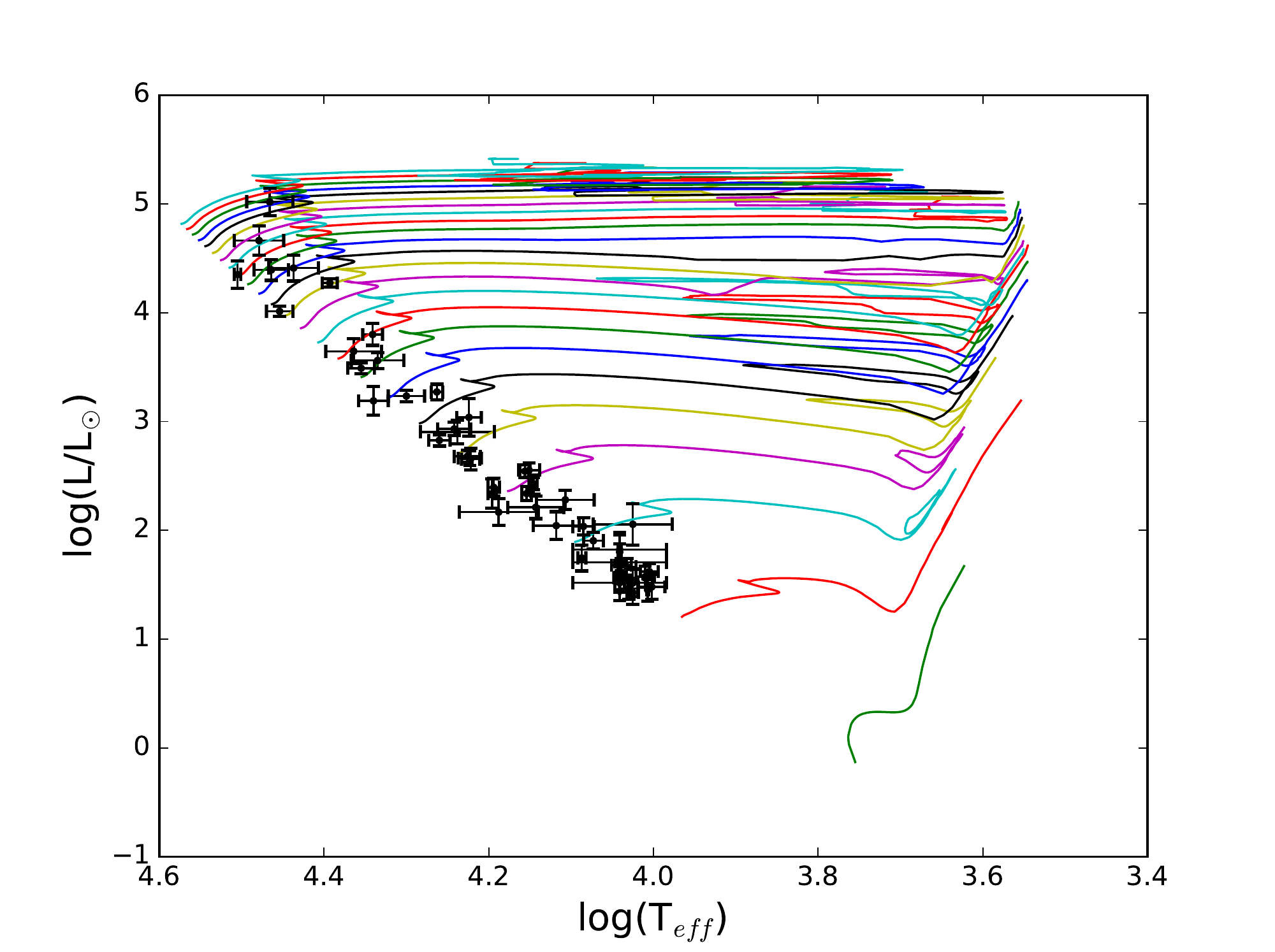}
    \label{fig:us_HR}
    \protect\caption{
    HR Diagram for Upper Scorpius includes stars from the most 
    massive down to \protect$\sim$2.3 \protect$M_{\odot}$.
    }
\end{figure}

The masses of all the B type members from each subgroup of Sco-Cen from the \citet{dezeeuw1999} study were estimated.  \autoref{fig:us_HR} shows the Hertzsprung-Russell (HR) diagram for the stars in the US subgroup.  Each HR diagram position was compared to the evolutionary tracks to estimate their individual masses which are listed in \autoref{tbl:USmass}.  This same process was repeated for UCL, and LCC in \autoref{tbl:UCLmass}, and \autoref{tbl:LCCmass}, respectively.  In the Sco-Cen subgroups, the uncertainty is listed for all masses above 5\,$M_{\odot}$.  Masses below this cutoff have a typical uncertainty of $\pm$0.2$M_{\odot}$.  The masses for Tuc-Hor members in \autoref{tbl:tuchormass} are adopted from the \citet{david2015} study with a typical uncertainty of approximately $\pm$0.2$M_{\odot}$.\par

\begin{table}
\centering
\caption{Masses for Upper Scorpius}
\label{tbl:USmass}
\setlength{\tabcolsep}{0.03in}
\begin{tabular}{llll}
\firsthline
Name          & Mass             & Name         & Mass          \\
              & (M$_{\odot}$)    &              & (M$_{\odot}$) \\
\hline
$\sigma$~Sco   & 21.0 $\pm$ 1.5  & HD 147701    & 4.0   \\
$\delta$~Sco   & 17.3 $\pm$ 1    & HD 147888    & 4.0   \\
$\tau$~Sco     & 16.0 $\pm$ 1    & HD 144661    & 3.8   \\
$\beta^1$~Sco  & 14.5 $\pm$ 0.5  & HD 144844    & 3.5   \\
$\pi$~Sco      & 14.0 $\pm$ 1    & HD 147932    & 3.5   \\
$\omega$~Sco   & 12.5 $\pm$ 0.5  & HD 142315    & 3.0   \\
HD 148184      & 12.0 $\pm$ 0.4  & HD 146285    & 3.0   \\
$\nu$~Sco      & 9.0  $\pm$ 0.5  & HD 147010    & 3.0   \\
HD 147933      & 9.0  $\pm$ 1    & HD 147196    & 3.0   \\
1~Sco          & 8.3  $\pm$ 0.2  & HD 149914    & 3.0   \\
$\rho$~Sco     & 8.0  $\pm$ 0.5  & HD 138343    & 2.8   \\
$\beta^2$~Sco  & 7.0  $\pm$ 0.5  & HD 143567    & 2.8   \\
13~Sco         & 6.8  $\pm$ 0.2  & HD 143600    & 2.5   \\
HD 142114      & 6.5  $\pm$ 0.5  & HD 143956    & 2.5   \\
HD 142301      & 5.8  $\pm$ 0.2  & HD 144586    & 2.5   \\
HD 142184      & 5.5  $\pm$ 0.2  & HD 145554    & 2.5   \\
HD 142378      & 5.5  $\pm$ 0.5  & HD 145631    & 2.5   \\
HD 142983      & 5.5  $\pm$ 0.5  & HD 145964    & 2.5   \\
HD 142990      & 5.5  $\pm$ 0.2  & HD 146331    & 2.5   \\
HD 142883         & 4.8          & HD 146416    & 2.5 \\
HD 144334         & 4.8          & HD 146706    & 2.5 \\
HD 139160         & 4.0          & HD 147553    & 2.5 \\
HD 142165         & 4.0          & HD 147648    & 2.5 \\
HD 142250         & 4.0          & HD 139486    & 2.4 \\
HD 145792         & 4.0          & HD 144569    & 2.3 \\
HD 146001         & 4.0          & HD 147955    & 2.3 \\
\hline
\end{tabular}
\begin{flushleft}
    \begin{footnotesize}The masses listed in this table are the masses above 2.3$M_{\odot}$. The masses that are below 5$M_{\odot}$ have a typical uncertainty of $\pm$0.2 $M_{\odot}$.
    \end{footnotesize}
\end{flushleft}
\end{table}
\begin{table}
\centering
\caption{Masses for Upper Centaurus-Lupus}
\label{tbl:UCLmass}
\setlength{\tabcolsep}{0.03in}
\begin{tabular}{llll}
\hline
Name            & Mass          & Name       & Mass          \\
                & (M$_{\odot}$) &            & (M$_{\odot}$) \\
\hline
$\delta$~Lup    & 13.0 $\pm$ 1.0     & HD 128819  & 3.9 \\
$\mu$~Cen       & 13.0 $\pm$ 1.0     & HD 147001  & 3.9 \\
$\alpha$~Lup    & 11.4 $\pm$ 0.3 & HD 142256  & 3.8 \\
$\mu^1$~Sco     & 10.0 $\pm$ 1.0 & HD 128207  & 3.7 \\
$\nu$~Cen       & 10.0 $\pm$ 1.0 & HD 136347  & 3.7 \\
$\beta$~Lup     & 9.5 $\pm$ 0.3 & HD 126548  & 3.6 \\
$\eta$~Cen      & 9.0 $\pm$ 0.2 & HD 138221  & 3.6 \\
$\mu^2$~Sco     & 9.0 $\pm$ 1.0 & HD 147628  & 3.6 \\
$\gamma$~Lup    & 9.0 $\pm$ 1.0 & HD 133937  & 3.5 \\
$\phi$~Cen      & 8.5 $\pm$ 0.5 & HD 140784  & 3.5 \\
$\kappa$~Cen    & 8.5 $\pm$ 0.5 & HD 124620  & 3.4 \\
$\eta$~Lup      & 8.5 $\pm$ 0.5 & HD 128775  & 3.3 \\    
HR 6143         & 8.0 $\pm$ 0.2 & HD 133652  & 3.3 \\
$\upsilon$~1Cen & 7.5 $\pm$ 0.5 & HD 135174  & 3.3 \\
$\epsilon$~Lup  & 7.5 $\pm$ 0.5 & HD 132238  & 3.2 \\
HD 133955       & 7.1 $\pm$ 1.0 & HD 140285  & 3.2 \\
$\chi$~Cen      & 7.0 $\pm$ 0.3 & HD 123445  & 3.1 \\
HR 5378         & 6.2 $\pm$ 0.2 & HD 124961  & 3.1 \\     
$\tau$~Lib      & 6.0 $\pm$ 0.2 & HD 121190  & 3.0   \\
$\theta$~Lup    & 5.8 $\pm$ 0.3 & HD 126475  & 3.0   \\
HR 5471         & 5.8 $\pm$ 0.2 & HD 126759  & 3.0   \\
HD 149711       & 5.8 $\pm$ 0.2 & HD 133880  & 3.0   \\    
HD 136664       & 5.7 $\pm$ 0.2 & HD 134837  & 3.0   \\     
HD 124367       & 5.5 $\pm$ 0.1 & HD 138923  & 2.8 \\
HD 133242       & 5.5 $\pm$ 0.2 & HD 143473  & 2.8 \\
HD 134687       & 5.5 $\pm$ 0.1 & HD 149425  & 2.8 \\ 
HD 138769       & 5.5 $\pm$ 0.1 & HD 135454  & 2.7 \\
HD 130807       & 5.1 $\pm$ 0.1 & HD 136482  & 2.7 \\
HD 150742       & 5.1 $\pm$ 0.1 & HD 143927  & 2.7 \\
HD 151109       & 5.0 $\pm$ 0.1 & HD 137919  & 2.6 \\
HD 128345       & 4.8           & HD 141327  & 2.5 \\
HD 143699       & 4.7           & HD 143022  & 2.5 \\
HD 140008       & 4.5           & HD 143939  & 2.5 \\
HD 137432       & 4.4           & HD 145880  & 2.5 \\
HD 131120       & 4.3           & HD 144591  & 2.4 \\
HD 150591       & 4.2           & HD 132094  & 2.4 \\
HD 135876       & 4.0           & HD 140840  & 2.3 \\
HD 147152       & 4.0           & HD 151726  & 2.3 \\
HD 126135       & 3.9             \\ 
\hline
\end{tabular}
\begin{flushleft}
    \begin{footnotesize}The masses listed in this table are the masses above 2.3$M_{\odot}$. The masses that are below 5$M_{\odot}$ have a typical uncertainty of $\pm$0.2 $M_{\odot}$.
    \end{footnotesize}
\end{flushleft}
\end{table}

\begin{table}
\centering
\caption{Masses for Lower Centaurus-Crux}\label{tbl:LCCmass}
\setlength{\tabcolsep}{0.03in}
\begin{tabular}{llll}
\hline
Name            & Mass          & Name       & Mass         \\
                & (M$_{\odot}$) &            & (M$_{\odot}$) \\
\hline
$\beta$~Cen     & 19.0 $\pm$ 1.0 & HD 107696       & 3.5 \\
$\alpha^1$~Cru  & 16.0 $\pm$ 1.0 & HD 112409       & 3.5 \\
$\beta$~Cru     & 14.5 $\pm$ 0.5 & HD 118354       & 3.3 \\
$\delta$~Cru    & 10.0 $\pm$ 1.0 & HD 114365       & 3.0 \\
$\alpha$~Mus    & 8.8  $\pm$ 0.3 & HD 95324        & 3.0 \\
HD 110879       & 7.1  $\pm$ 0.3 & HD 104600       & 2.9 \\
$\xi^2$~Cen     & 7.0  $\pm$ 0.3 & HD 110506       & 2.9 \\
$\mu^1$~Cru     & 7.0  $\pm$ 0.3 & HD 113902       & 2.9 \\
$\sigma$~Cen    & 7.0  $\pm$ 0.4 & HD 110020       & 2.8 \\
$\zeta$~Cru     & 5.8  $\pm$ 0.2 & HD 110461       & 2.8 \\
HR 4618         & 5.7  $\pm$ 0.2 & HD 104080       & 2.7 \\
HD 108257       & 5.4  $\pm$ 0.4 & HD 114772       & 2.6 \\
HD 98718        & 5.1  $\pm$ 0.1 & HD 100546       & 2.5 \\
HD 110956       & 5.0  $\pm$ 0.1 & HD 104839       & 2.5 \\
HD 112091       & 5.0  $\pm$ 0.1 & HD 107301       & 2.5 \\
HD 112078       & 4.9            & HD 109195       & 2.5 \\
HD 113703       & 4.8            & HD 110737       & 2.5 \\
HD 116087       & 4.8            & HD 112381       & 2.5 \\
HD 100841       & 4.5            & HD 115470       & 2.5 \\
HD 103079       & 4.5            & HD 115583       & 2.5 \\
HD 90264        & 4.3            & HD 115988       & 2.5 \\
HD 115823       & 4.0            & HD 119419       & 2.5 \\
HD 114529       & 3.8            & HD 117484       & 2.3 \\
HD 114911       & 3.8            & HD 118697       & 2.3 \\
\hline
\end{tabular}
\begin{flushleft}
    \begin{footnotesize}The masses listed in this table are the masses above 2.3$M_{\odot}$. The masses that are below 5$M_{\odot}$ have a typical uncertainty of $\pm$0.2 $M_{\odot}$.
    \end{footnotesize}
\end{flushleft}
\end{table}
\begin{table}
\centering
\caption{Masses for Tucana-Horologium}\label{tbl:tuchormass}
\setlength{\tabcolsep}{0.03in}
\begin{tabular}{ll}
\hline
Name          & Mass          \\
              & (M$_{\odot}$) \\
\hline
HIP 100751    & 5.35 $\pm$ 0.40\\
HD 14228      & 3.49 $\pm$ 0.15\\
HIP 2484      & 2.67 $\pm$ 0.15\\
HIP 12394     & 2.53 $\pm$ 0.15\\
HIP 118121    & 2.31 $\pm$ 0.15\\
HIP 2578      & 2.11 $\pm$ 0.10\\
HIP 2487      & 1.75 $\pm$ 0.10\\
HIP 104308    & 1.58 $\pm$ 0.10\\
\hline
\end{tabular}
\begin{flushleft}
    \begin{footnotesize}The masses listed in this table are the stars above 1.5 M$_{\odot}$ that were adopted from the \citet{david2015} study.
    \end{footnotesize}
\end{flushleft}
\end{table}

\subsection{Black Holes}
Observations of SNe events from the past 15 years and theoretical stellar models have strongly suggested that SNe progenitors exist within a range between 8\,M$_{\odot}$ to $\simeq$18\,M$_{\odot}$, where anything more massive than $\simeq$18\,M$_{\odot}$ will not become a SN, but instead will collapse into a black hole \citep{smartt2015}. The support for this upper limit was strengthened by the recent discovery of a $\sim$25\,M$_{\odot}$ red supergiant that failed to SN \citep{adams2017}. This evidence explains the discrepancy between observed and expected SNe events within the solar neighborhood.  Using $\simeq$18\,M$_{\odot}$ as an upper limit will provide better constraints on each subgroup being considered in this project for the SN evidence discussed in the \citet{wallner2016} study.  \par

\subsection{Ocean Core Samples and Travel Time}
A recent examination of several ocean core samples by \citet{wallner2016} found radionuclide signals in the core samples from different locations on Earth.  The \citet{wallner2016} samples contained radionuclides, likely from SN ejecta corresponding to two distinct events.  One event, which is referred to here as the `Recent Event', was likely from SN ejecta that landed 2-3 Myr ago on Earth.  The other event, referred to here as the `Older Event', was likely from SN ejecta that landed on Earth 7-9 Myr ago.  This present work considers the origin of these two events as core-collapse SNe, and attempts to determine the likely birth-site of these events, taking into account an assumed initial mass function, the masses and possible progenitor birth-sites of nearby young associations, and a timeline that is consistent with the assumed mean ages of these associations.  \par

The uncertainty in the ocean core samples is due to the natural contamination of oceanic isobars \citep{fields2005}. The $^{60}$Fe could not have had a terrestrial origin because the signal had been found in every ocean, leading to the assumption that there is a uniform distribution over the Earth \citep{wallner2016}.  \par 

The travel time adopted for this project is a rounded estimate of $\sim$0.18-0.25 Myr from \citealt{feige2014} and 0.14-0.98 Myr from \citealt{fry2015}.  This current work will assume a mean travel time of $\sim$0.5\,Myr for the Sco-Cen subgroups and Tuc-Hor, considering the difference in calculations.  These travel times are less than the uncertainties in the timeline of the $^{60}$Fe signal. \par

\section{Analysis}
\begin{figure}[h!]
    \centering
    \includegraphics[width=\columnwidth]{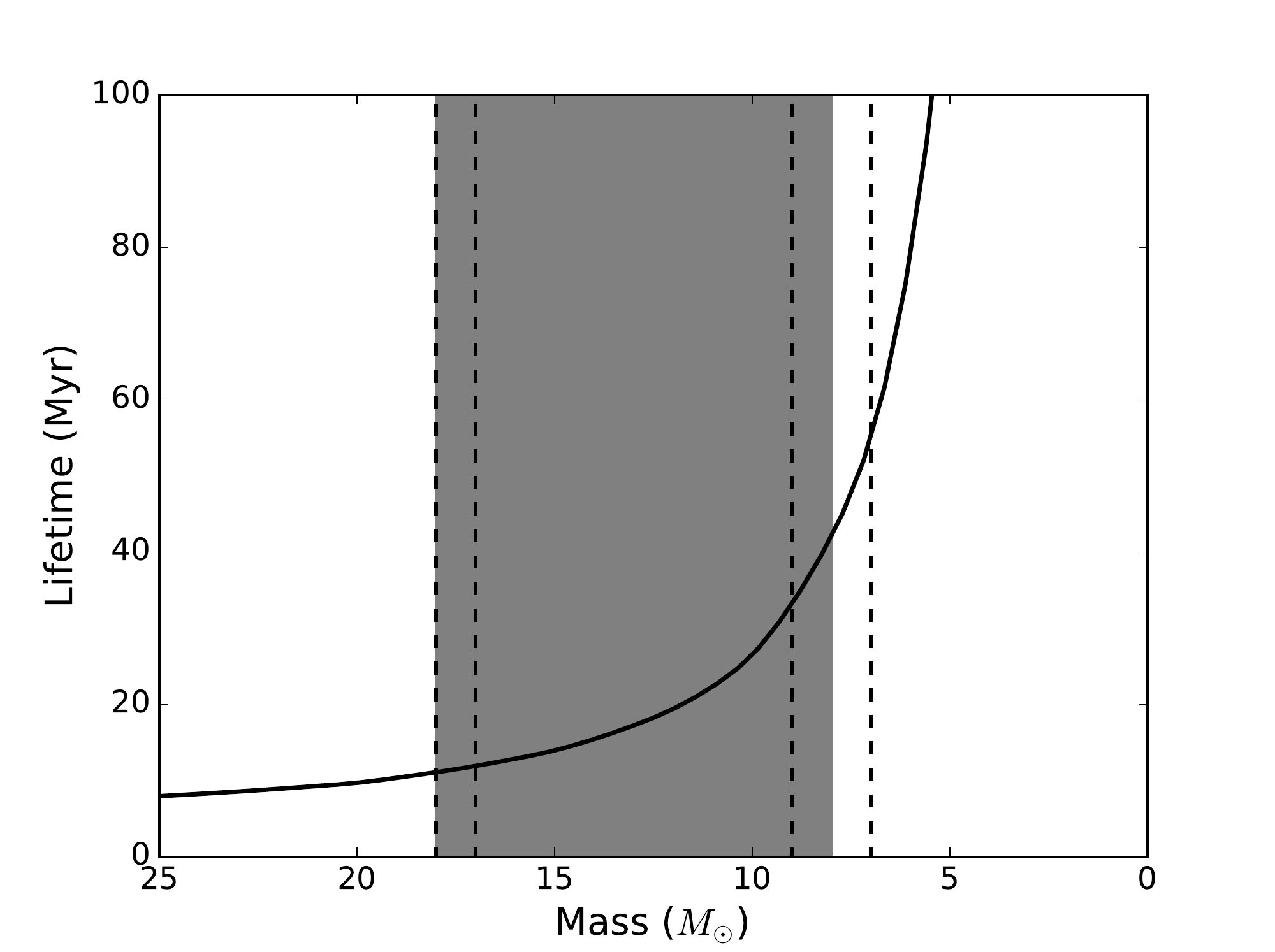}
    \caption{
    Mass-Lifetime Relationship 
    These stellar lifetimes include a rotational velocity of $V/V_{crit}$ = 0.355 and a solar metallicity of Z = 0.014 that were interpolated from the Ekstr{\"o}m evolutionary tracks.  The stellar masses were obtained from the HR diagrams.  The dashed lines indicate the uncertainties associated with the adopted upper and lower limits for the masses of SNe candidates.
    }
    \label{fig:age_mass}
\end{figure}

It was decided to investigate both the random and optimal sampling initial mass function (IMF) techniques to compare the differences in results for each group of stars.  The expected lifetime of each mass ranging from 1-70$M_{\odot}$ were interpolated from the \citet{ekstrom2012} evolutionary tracks and was used to generate a mass-lifetime relationship.  An upper mass limit of 70$M_{\odot}$ was chosen, since it was more than 3 times larger than the largest mass still present in Sco-Cen.  The mass-lifetime relationship is shown in \autoref{fig:age_mass}.  \citet{preibisch2002} have examined the IMF of US and found that a slope of -2.35 accurately describes Sco-Cen, so this slope was adopted for both random and optimal sampling.\par
The IMF is a mathematical model used to describe the distribution of stars born from the same event.  Being born from the same event, the stars within an association will be approximately `coeval', which provides an advantage in developing a model for the IMF.  The IMF is expressed as a power-law whose slope determines the distribution of stellar masses \citep{kroupa2013}.  Even though the IMF of a population cannot be directly measured, great progress has been made in determining the slopes for the IMF depending on various characteristics of these populations.  These characteristics are inferred from the HR diagrams.  Understanding the IMF will lead to a better interpretation of the formation of stars \citep{bastian2010}.  \par

The IMF is described in the following function:

\begin{equation}
\xi(m) \propto m^{-\alpha} \label{eqn:imf1}
\end{equation}

where $\xi(m)\,dm$ is the number of stars within an association on a mass interval [$m, m + dm$].  Salpeter's IMF ($\alpha$ = 2.35) has been shown to describe the masses in a cluster that are greater than 1\,$M_{\odot}$ \citep{salpeter1955}, and appears to be universal among young clusters and associations as well as older globular clusters \citep{bastian2010}.  The IMF is typically used as a multiple power-law distribution using different slopes for masses smaller than 1 $M_{\odot}$.  However, this multiple power-law will not be necessary because the distribution of SN-eligible stars is constrained within the limits of only one of these power-laws, $\alpha$ = 2.35 for $M{\textgreater}1M_{\odot}$ stars.  \par

\subsection{Random Sampling}
Two different approaches for the IMF will be explored here:  random and optimal sampling.  Random sampling involves producing a synthetic population of stars that are independent of each other and can be performed with or without constraints \citep{kroupa2013}.  These synthetic populations are created over a prior distribution that corresponds to the spread of stellar masses in a system.  The probability density function is found between the minimum and maximum values (no constraints using the physical mass limit $m_{max}$ = $\infty$).  This probability density is described as a power-law distribution from which thousands of populations of stars can be generated in order to determine the most likely shape of the cluster over a given $\alpha$ \citep{kroupa2013}.  \par 

The normalization constant for US for the proportionality in \autoref{eqn:imf1} was calculated by using the number of stars above 2.3\,$M_{\odot}$ (see \autoref{fig:us_HR}).  The list of the stellar masses for US, which were inferred from their HR Diagram, are listed in \autoref{tbl:USmass}.  Using the mass-lifetime relationship from \autoref{fig:age_mass}, a synthetic population of 25,000 stars was developed for each group with their corresponding lifetimes.  Both the normalization and Salpeter's constants were used to create an extrapolative model for the missing SN progenitor.  Please note that these simulations do not include any additional assumptions such as ejecta travel time or time of SN event.  The simulations only include the data from their respective subgroups (i.e. rotational velocity, metallicity, and mass) and are not sensitive to their assumed ages.  Additional characteristics (e.g., subgroup age, black holes, runaway stars, and travel time) will be discussed in greater detail in Section \ref{sec:results}.\par

\subsection{Estimating Missing Mass}
Instead of choosing masses at random, optimal sampling provides stellar masses that are perfectly distributed according to the IMF that chooses masses in descending order from $m_{max}$ \citep{kroupa2013}.  Optimal sampling is solved iteratively by first finding the normalization constant then the most massive star \citep{kroupa2013} by the following two equations:
\begin{equation}1 = \int_{m_{i}}^{m_{max}} \xi(m)dm \label{eqn:opt_norm} \end{equation}
\begin{equation}M_{encl}(m_{i}) - m_{i} = \int_{m_{i}}^{m_{i+1}}\xi(m)dm \label{eqn:opt_massive} \end{equation}
where the term $M_{encl}$ is the enclosed mass of the entire cluster or subgroup of stars and $M_{encl}(m_{i})$ is the correction term since this star must remain between $m_{i}$ and $m_{max}$ \citep{kroupa2013}. \par 

Though optimal sampling was not used for this project, it helped inspire a process to estimate the next most massive star of each group, referred to in this work as the Estimating Missing Mass Function (EMMF).  After finding the normalization constant from \autoref{eqn:opt_norm} by counting the stars in a range of masses from each group, a modified version of \autoref{eqn:opt_massive} was used.  Instead of solving the equation iteratively in descending order, everything--including the normalization constant--was put back into \autoref{eqn:opt_norm} with the most massive cataloged star ($m_{current}$) as the lower limit and the unknown variable to solve being the upper limit in the equation.  This modified equation is represented by the following expression:
\begin{equation}1 = \int_{m_{current}}^{m_{unknown}} \xi(m)dm \label{eqn:missing_mass} \end{equation}
where $m_{unknown}$ is the unknown variable and the mass of the calculated missing star. Please note that the EMMF used the same data as Random Sampling and is independent of the ages of their respective subgroups.\par  

\begin{table}[h!]
\centering
\caption{Possible Progenitor Mass For One Event}\label{tbl:optsample1}
\setlength{\tabcolsep}{0.03in}
\begin{tabular}{lll}
\hline
Subgroup              & Missing Mass   &    Lifetime \\
                      & ($M_{\odot}$)  &    of Mass (Myr)\\
\hline
Lower Centaurus-Crux  &  25.4$\pm$1.4  &    7.5-8.2 \\
Tucana-Horologium     &  9.36$\pm$3.2  &    19.4-79.0\\
Upper Centaurus-Lupus & 14.2$\pm$0.2   &    15.4\\
Upper Scorpius        & 28.8$\pm$1.7   &    6.9-7.5\\
\hline
\end{tabular}
\begin{flushleft}
    \begin{footnotesize}The uncertainties in the masses given represent a 95\% confidence interval.
    \end{footnotesize}
\end{flushleft}
\end{table}
\begin{table}[h!]
\centering
\caption{Possible Progenitor Mass For Two Events}\label{tbl:optsample2}
\setlength{\tabcolsep}{0.03in}
\begin{tabular}{lll}
\hline
Subgroup    & Missing Mass    &Lifetime \\
    &   ($M_{\odot}$)   &   of Mass (Myr)\\
\hline
Lower Centaurus-Crux &  42.1$\pm$4.9 &  5.2-6.0 \\
Tucana-Horologium    & ------------  & ------------\\
Upper Centaurus-Lupus    &  15.8$\pm0.2$ &  12.9  \\
Upper Scorpius  &   51.7$\pm7.1$  & 4.5-5.3 \\
\hline
\end{tabular}
\begin{flushleft}
    \begin{footnotesize}The uncertainties in the masses given represent a 95\% confidence interval. The total mass of Tuc-Hor is too small to provide a possible progenitor to be considered for more than one event.
    \end{footnotesize}
\end{flushleft}
\end{table}

To check the validity of this equation, this process was used against the present-day stars (each time adopting the new normalization constant produced from \autoref{eqn:opt_norm}) which verified the accuracy of this method.  \autoref{tbl:optsample1} and \autoref{tbl:optsample2} provide the estimated missing masses from each group using this same method.  This process is equivalent to placing each mass into its own bin, as discussed in \citet{maiz2005}. Furthermore, the results in \citet{maiz2005} indicate that even when the bin sizes are very small, even as low as one star per variable-sized mass bin, the resulting biases in recovering the IMF are very small.  This implies that the EMMF method described here is also subject to very little bias in determining the upper limit of the mass bin, and thus very little bias in inferring the mass of the missing supernova progenitors in the associations examined here.\par

\section{Results}\label{sec:results}
\subsection{Random Sampling}
\begin{figure}
    \centering
    \includegraphics[width=\columnwidth]{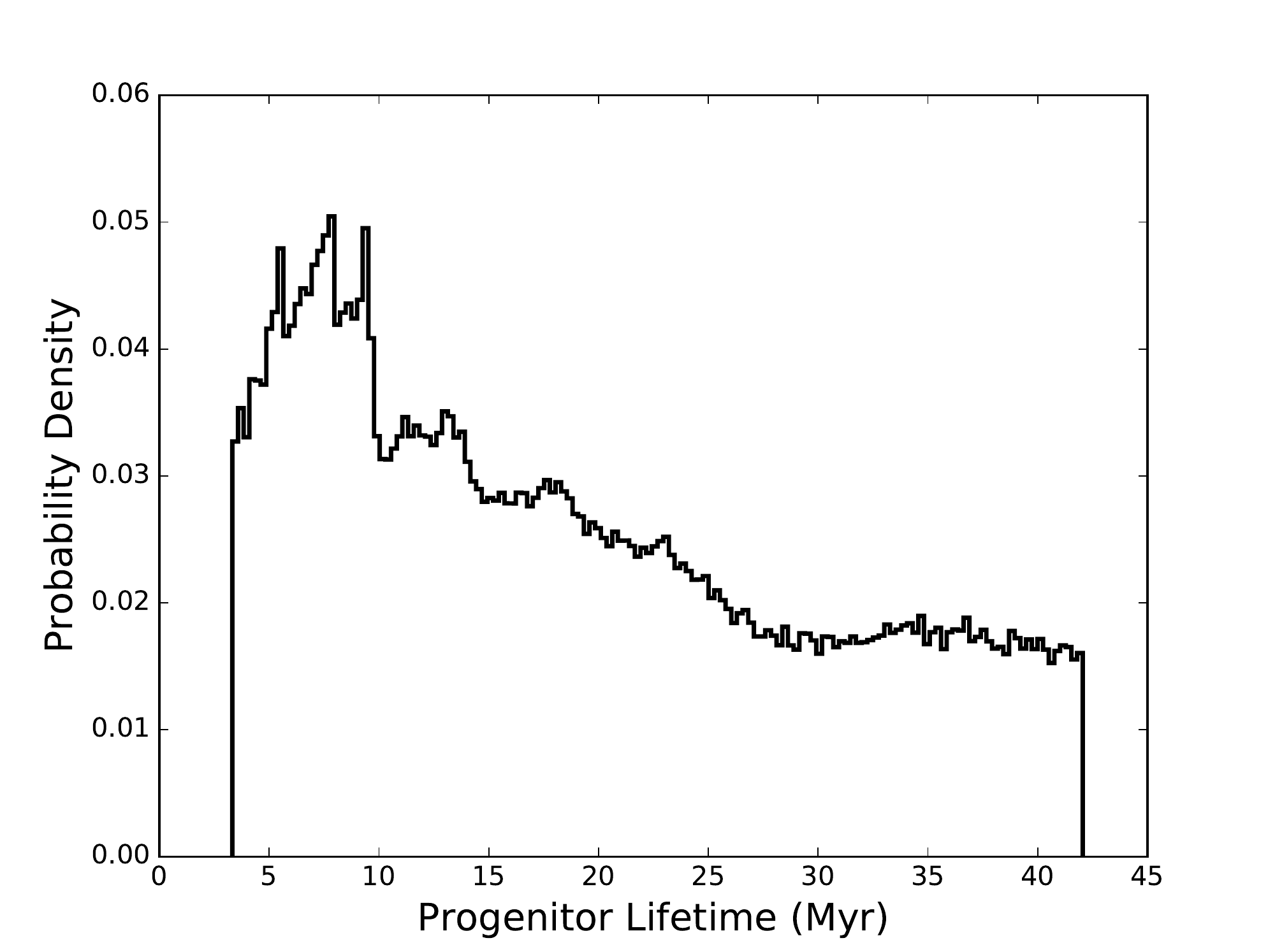}
    \caption{
    Age Histogram of Upper Scorpius. 
    These ages were drawn from a synthetic population of 25,000 stars that were modeled from the mass-lifetime relationship discussed in 3.1 and are the results of the random sampling method.
    }
    \label{fig:us_hist}
\end{figure}

The masses of each star were inferred from their HR diagrams for each subgroup considered for this project.  These masses were used to establish a mass-lifetime relationship by using the models from the \citet{ekstrom2012} evolutionary tracks (refer to \autoref{fig:age_mass}).  This mass-lifetime relationship was used to translate the results of random sampling (refer to \autoref{fig:us_hist}) into a lifetime for the possible missing mass from each subgroup within the SN-eligible range $\sim$8\,M$_{\odot}$ to $\simeq$18\,M$_{\odot}$.  One interesting discovery from the random sampling method was that each subgroup had near-identical results though they had widely varying normalization constants because of the differences in the range of masses in relation to the count of masses in each group.  Since the results for the remaining subgroups were so similar to the results for US, they will not be included in this paper.\par
\subsection{Estimating Missing Mass}
Instead of using the IMF in the traditional fashion by solving for each mass iteratively and in descending order, the possible SN progenitor was solved in ascending order by maintaining the normalization and Salpeter's constants with the most massive cataloged star being the lower limit.  The results for each subgroup can be found in \autoref{tbl:optsample1} and \autoref{tbl:optsample2}.  Using both of these tables, LCC and US were eliminated as candidate birth-sites because they had large masses outside of the 8\,M$_{\odot}$ to $\simeq$18\,M$_{\odot}$ range, a range supported by the recent discovery of a $\sim$25 M${_\odot}$ red supergiant which failed to SN \citep{adams2017}.  Tuc-Hor, in comparison, could only have been responsible for one of the two events mentioned in the \citet{wallner2016} study.  Since the missing mass for Tuc-Hor has an uncertainty of approximately 3\,M$_{\odot}$, there is a possibility that Tuc-Hor could be responsible for either the Recent Event (2-3 Myr) or the Older (7-9 Myr) SN Event.  \par

Up to this point, the ages of the associations have not been considered since they were not relevant to determining the masses of the SN progenitors.  When considering the likely birth site for the progenitor, the lifetime of the proposed progenitor must be consistent with published mean ages of each association.  Published ages for US range from 5-11 Myr \citep{preibisch1999, pecaut2012, kraus2015, david2015, feiden2016}, while ages for UCL and LCC range from 10-20 Myr \citep{mamajek2002, sartori2003, pecaut2012, song2012}. The ages that will be adopted for US, UCL, and LCC are 10$\pm$3\,Myr, 16$\pm$2\,Myr, and 15$\pm$3\,Myr, respectively \citep{pecaut2016}.  The adopted age for Tuc-Hor will be 45$\pm$4 Myr \citep{bell2015}. \par

As the predicted progenitor from Tuc-Hor has such a large mass range, the mass range will be evaluated in sections (6-7, 8, 9, and 10-12\,$M_{\odot}$) in order to eliminate or verify the possibility that either SN event took place in Tuc-Hor.  The 6-7\,M$_{\odot}$ range yields a lifetime between $\sim$55-80\,Myr which is inconsistent with the age of Tuc-Hor and the time of the SN, as well as being inconsistent with the SN lower limit.  If the progenitor was 8\,M$_{\odot}$ (42\,Myr lifetime) and factoring in  $\sim$0.5\,Myr travel time, then the lifetime of the progenitor would fit within the timeline of the Recent Event.  In comparison, if the progenitor was 9\,M$_{\odot}$ (33\,Myr lifetime), then the age of the progenitor could only be eligible for the Older Event.  If the progenitor was 10\,M$_{\odot}$ (26\,Myr lifetime) and factoring in $\sim$0.5\,Myr travel time, then the lifetime of the progenitor would be too young for either event, thus eliminating the 10-12\,M$_{\odot}$ range. \par 


For UCL, the predicted progenitor masses from \autoref{tbl:optsample1} and \autoref{tbl:optsample2} will be used in the same manner as Tuc-Hor--against its mean age.  If UCL was responsible for one of the two events, having a progenitor mass of $\sim$14\,M$_{\odot}$ and a lifetime of $\sim$15\,Myr would be inconsistent for the Older SN Event but within the lifetime range for the Recent SN Event.  The same result will also be true for UCL if it was responsible for two events.  Being responsible for two events required a progenitor mass of 16\,M$_{\odot}$ ($\sim$13\,Myr), again, being consistent with only the Recent Event.  The results for UCL are consistent with the time frame for a SN event that was outlined by \citealt{feige2010}, \citealt{feige2017}, and \citealt{sorensen2017}. \par

Since this method depends on the total stellar mass of the association, the occurrence of runaway stars may affect the results of the EMMF. The number of runaway stars within a cluster appears to vary between $\sim$10 \citep{blaauw1961} to $\sim$90$\%$ \citep{dewit2005}, or between 2 to 27$\%$ when those estimates are combined \citep{lamb2008}. Earlier works suggested that 6 stars (HIP 42038, HIP 46950, HIP 48943, HIP 69491, HIP 76013, and HIP 82868) may have originated from Sco-Cen, but were later found to have come from  IC 2391 and IC 2602 \citep{jilinski2010}. Another study found that the OB density in US has not changed significantly \citep{pflamm2006}.   \par

Since this issue continues to be a topic of discussion and is not a main theme of this paper, it was decided that the calculations for the EMMF method should consider an additional 3 missing stars from each subgroup.  After repeating the analysis, it was found that the original results displayed in \autoref{tbl:optsample1} and \autoref{tbl:optsample2} would not be significantly altered since the normalization constants were not drastically changed.  Assuming 3 ejected massive stars in each association, the mass of progenitors of a single event in each group are as follows: LCC (24.9$\pm$1.2\,M$_{\odot}$), UCL (14.2$\pm$0.2\,M$_{\odot}$), US (28.2$\pm$1.5\,M$_{\odot}$), and Tuc-Hor (7.5$\pm$1.1\,M$_{\odot}$); the results for two events are:  LCC (38.4$\pm$3.4\,M$_{\odot}$), UCL (15.7$\pm$0.2\,M$_{\odot}$), and US (46.5$\pm$4.9\,M$_{\odot}$). Since the addition of potential runaway stars had little effect on the predicted SNe candidates from each subgroup, it was decided to maintain the original results displayed in \autoref{tbl:optsample1} and \autoref{tbl:optsample2}.  Please note that the results accounting for the runaway stars live within the uncertainties found in the original results. \par

\subsection{Comparison of Previous Work}

\begin{table}
\centering
\caption{Comparison of Results from Previous Work for Lower Centaurus-Crux}
\label{tbl:BLCC}
\setlength{\tabcolsep}{0.03in}
\begin{tabular}{lllll}
\hline   
SN & Lifetime & Time SN & Implied & Consistent \\
Progenitor & & Occurred & Age for & with Subgroup \\
Mass (M$_{\odot}$) & (Myr) & (Myr) & LCC (Myr) & Age?   \\
\hline
18.61 & Blackhole &------------ &------------\\
15.36 & 14 & 10.0 & 24 & Yes*\\
13.12 & 17 & 8.0 & 25 & Yes*\\
11.48 & 19 & 6.1 & 25.1 & Yes*\\
10.21 & 26 & 4.2 & 30.2 & No\\
9.21  & 33 & 2.3 & 35.3 & No\\
\hline
\end{tabular}
\begin{flushleft}
    \begin{footnotesize}`SN Progenitor Mass' and `Time SN Occurred' are the results of the supernova progenitor mass and the time the supernova occurred as established by the \citet{breitschwerdt2016} study.  The `Lifetime' column contains the approximate lifetimes of the `SN Mass' that were interpolated from the \citet{ekstrom2012} study.  `Implied Age' is the age implied for the association from the putative progenitor mass and the mass-lifetime relationship.  See Section 4.3 for discussion. Yes* indicates consistency with some regions of the subgroup, based on the age map contained in \citet{pecaut2016}
    \end{footnotesize}
\end{flushleft}
\end{table}

\begin{table}
\centering
\caption{Comparison of Results from Previous Work for Upper Centaurus-Lupus}
\label{tbl:BUCL}
\setlength{\tabcolsep}{0.03in}
\begin{tabular}{lllll}
\hline   
SN & Lifetime & Time SN & Implied & Consistent \\
Progenitor & & Occurred & Age for & with Subgroup \\
Mass (M$_{\odot}$) & (Myr) & (Myr) & UCL (Myr) & Age?   \\
\hline
19.86 & Blackhole &------------ &------------\\
17.34 & 12 & 11.3 & 23.3 & Yes*\\
15.41 & 14 & 10.0 & 24 & Yes*\\
13.89 & 15 & 8.7 & 23.7 & Yes*\\
12.65 & 17 & 7.5 & 24.5 & Yes*\\
11.62 & 20 & 6.3 & 26.3 & No\\
10.76 & 22 & 5.0 & 27 & No\\
10.02 & 26 & 3.8 & 34.8 & No\\
9.37  & 33 & 2.6 & 35.6 & No\\
8.81  & 33 & 1.5 & 34.5 & No\\
\hline
\end{tabular}
\begin{flushleft}
    \begin{footnotesize}`SN Progenitor Mass' and `Time SN Occurred' are the results of the supernova progenitor mass and the time the supernova occurred as established by the \citet{breitschwerdt2016} study.  The `Lifetime' column contains the approximate lifetimes of the `SN Mass' that were interpolated from the \citet{ekstrom2012} study.  `Implied Age' is the age implied for the association from the putative progenitor mass and the mass-lifetime relationship.  See Section 4.3 for discussion.  Yes* indicates consistency with some
    regions of the subgroup, based on the age map contained
    in \citet{pecaut2016}
    \end{footnotesize}
\end{flushleft}
\end{table}

Previous work in this field by \citet{breitschwerdt2016} established results that conflict with the findings of this project and will be addressed by each subgroup (UCL and LCC). In the \citet{breitschwerdt2016} study, they used the $^{60}$Fe from the Recent Event to establish a time-line of the formation of the Local Bubble by predicting the frequency and the mass of multiple SNe events.  In both \autoref{tbl:BLCC} and \autoref{tbl:BUCL}, the `SN Mass' and `Time SN Occurred' columns contain the masses of the progenitors and the time they exploded, respectively, as determined by \citet{breitschwerdt2016}.  The `Lifetime' column contains the rounded lifetimes of the `SN Mass' as established by the Ekstr{\"o}m evolutionary tracks.  The remaining column in both tables is the result of adding the data from the `Lifetime' and the `Time SN Occurred' columns which produce the expected ages for each respective subgroup.\par

In both \autoref{tbl:BLCC} and \autoref{tbl:BUCL}, the mass in the first row exceeded the upper limit of $\simeq$18$M_{\odot}$ set by \citet{smartt2015}, so these masses were disregarded. Comparing the results from \autoref{tbl:BLCC} and the adopted mean age for LCC, 15$\pm$3\,Myr \citep{pecaut2016}, shows that the masses predicted by \citet{breitschwerdt2016} are incompatible with this subgroup. An additional comparison between the masses from \autoref{tbl:BLCC} and \autoref{tbl:LCCmass} shows that the \citet{breitschwerdt2016} study concluded that 5 of the masses to SN are smaller than the largest masses currently in LCC (refer to \autoref{fig:age_mass} to review the mass-lifetime relation).  The mass-lifetime relation indicates that the smaller a mass is, the longer it will live--meaning that it is highly improbable for masses with a significantly longer lifespan to SN before the larger masses.  However, the significant age spread indicated by the age map in \citet{pecaut2016} indicates that the masses within the 12-27 Myr range are plausible candidates for LCC. \par 

The adopted mean age for UCL is 16$\pm$2\,Myr \citep{pecaut2016}, so the masses and their respective times of death within \autoref{tbl:BUCL} imply that it is inconsistent with the mean age.  Much like LCC, the \citet{breitschwerdt2016} study concluded that there were masses smaller than those currently in UCL to have already exploded as a SN. However, there is a significant age spread within this subgroup where portions are $\sim$24\,Myr old \citep{pecaut2016}.  Accepting the masses between the 14-24\,Myr range would eliminate the last 5 masses from \autoref{tbl:BUCL} for eligibility as progenitor candidates. \par

Because various ages for the subgroups of Sco-Cen appear in the literature, this contribution will consider the implications of younger ages; $\sim$5\,Myr for US and $\sim$10 Myr for UCL and LCC \citep{preibisch2002,song2012}.  If the true age of Sco-Cen were younger, then the predicted SN progenitors for the Sco-Cen subgroups found in \autoref{tbl:optsample1} and \autoref{tbl:optsample2} would be incompatible with their respective ages.  Similarly, the results obtained from the \citet{breitschwerdt2016} study would also be incompatible since the lifetimes of the SNe progenitors would be more than double the mean subgroup age. \par

In addition to considering a younger age for Sco-Cen, it would also be important to recognize that an 18\,M$_{\odot}$ upper limit for a SN may not be a definite cutoff since the mass from the \citet{adams2017} study was $\sim$25 M$_{\odot}$.  In \autoref{tbl:BLCC}, the lifetime for a 18.61 M$_{\odot}$ progenitor would imply an age of 22.3\,Myr for LCC which would be consistent with the age map for LCC by \citet{pecaut2016}.  Likewise, the 19.86 M$_{\odot}$ progenitor for UCL in \autoref{tbl:BUCL} would imply an age of 22.3\,Myr and would also be consistent with the older $\sim$24 Myr portions of UCL found within the \citet{pecaut2016} age map. \par

\section{Discussion} \label{sec:discussion}
Though the results from random sampling were not used in this project, it is an important point of discussion.  The results from every subgroup analyzed produced near-identical results despite having different normalization constants.  One explanation for this phenomenon is that random sampling was not the appropriate method for this project.  Random sampling is a probability distribution which is the addition of many of the same results over the same distribution.  The results, therefore, were very similar because each subgroup had similar initial and cutoff values which produced very similar histograms. \par

Another important point of discussion is with the results obtained from the EMMF.  Each subgroup was treated as a whole, as opposed to sectioning off each subgroup into age-relevant groupings as indicated in the age map by \citet{pecaut2016}.  This method would perhaps give a better idea of where a SN may have occurred in comparison to the results of this project. These age spreads are one possible reason for producing such a large and unusable progenitor mass within US and LCC and is a prospect for future research within this field.\par

Some of the conclusions from previous work on this subject are inconsistent with the assumed ages of Sco-Cen, as addressed in Section \ref{sec:results}.  This was done by comparing the resulting masses from the \citet{breitschwerdt2016} study to the expected lifetimes from the Ekstr{\"o}m evolutionary tracks.  Another constraint for the masses from the \citet{breitschwerdt2016} study was provided by the upper limit for SNe by the \citet{smartt2015} study.  Once all inconsistencies were eliminated, only 4 possible masses within UCL and 3 within LCC remained within reason for the cause for the Recent (2-3\,Myr) Event.\par

\section{Conclusions}
The results from random sampling provided no new insights to the point of origin for the SNe events discussed in the \citet{wallner2016} study.  This was due to the vague results that were difficult to interpret but still consistent with the EMMF.  Initially, random sampling was to be used as a reference for comparison between each subgroup, but instead has become a reference for comparison between different sampling methods. \par

In contrast, the EMMF proved an excellent method for generating the missing star (the next largest mass) within a group which allowed for a straightforward interpretation of results to compare the SN timeline, the mass-lifetime relationship, and the adopted mean ages of each association to test each birth-site for plausibility.  This process had been adapted to predict current stellar masses within a very small range of uncertainty.  The only problem encountered with this method was that it was applied to the entire subgroup and not to age-relevant sections as described in \citet{pecaut2016}.\par

US and LCC were eliminated as possible points of origin for the SNe events based on the size of the masses from the EMMF which produced stellar masses that were beyond the accepted range indicated by \citet{smartt2015}.  Tuc-Hor, surprisingly, was found to fit within the time constraints for either event.  The problem with Tuc-Hor, though, is that the masses within the moving group are extremely small, meaning it could have only produced one SN.  UCL, in comparison, was found to only fit within the Recent Event based on its adopted mean age.  By process of elimination, Tuc-Hor is favored for the Older Event.

\acknowledgements
First, we would like to thank Rockhurst University for providing us with the Dean's Undergraduate Fellowship for Research and Creative Activity that launched this project.  Thank you to Cool Stars 19, University of Nebraska-Lincoln, University of Missouri-Kansas City, and Rockhurst University for allowing us to present the findings of our research to such a wide variety of audiences.  Thank you to Katie Boyce, Rachel Schaff, Grant Eckelkamp, and Skylar Smith for asking questions and making helpful suggestions.  And last, we would like to thank the referee for providing helpful comments and suggestions which improved the quality of the paper.
  
\bibliographystyle{an}
\bibliography{main}

\end{document}